\documentclass[11pt,twoside]{article}

\usepackage{asp2014}

\aspSuppressVolSlug
\resetcounters

\begin{document}

\title{Towards Robotic Operation with the First G-APD Cherenkov Telescope}

\author{Maximilian Noethe,$^1$ Dominik Neise,$^2$ and Sebastian A. Mueller$^2$, for the FACT Collaboration
\affil{$^1$TU Dortmund, Dortmund, Germany; \email{maximilian.noethe@tu-dortmund.de}}
\affil{$^2$ETH Zurich, Zurich, Switzerland}
}

\paperauthor{Maximilian~Noethe}{maximilian.noethe@tu-dortmund.de}{}{TU Dortmund}{Experimental Physics 5}{Dortmund}{NRW}{44227}{Germany}
\paperauthor{Dominik~Neise}{neised@phys.ethz.ch}{}{ETH Zurich}{Institute for Particle Physics}{Dortmund}{NRW}{44227}{Germany}
\paperauthor{Sebastian~Achim~Mueller}{sebmuell@phys.ethz.ch}{}{ETH Zurich}{Institute for Particle Physics}{Dortmund}{NRW}{44227}{Germany}

\begin{abstract}
  The First G-APD Cherenkov Telecope is an Imaging Air Cherenkov Telescope operating since 2011 at the Observatorio del Roque de los Muchachos.
  One of the major goals of the FACT collaboration is to achieve robotic operation of the telescope.
  Since 2011 FACT is operated remotely. 
  To reduce the necessity of human intervention, several programs were developed, 
  most notably the \texttt{shifthelper} together with the \texttt{pycustos} library.
  This software monitors the telescope system and environmental conditions and calls  the shifters in case human intervention is required. 
  This will lead to FACT being the first IACT with all shifters asleep during regular observations.
  The software presented here is open source and under MIT License.
\end{abstract}

\section{The First G-APD Cherenkov Telecope}
FACT, the First G-APD Cherenkov Telescope, is an Imaging Atmospheric Cherenkov Telescope (IACT) located at the Observatorio del Roque de los Muchachos on the Canary Island of La~Palma.~\citep{fact-reference}, \citep{fact-performance}

After starting operation in October 2011, the FACT collaboration pursued three major goals:
Continuous monitoring of the brightest gamma-ray sources,
pioneering the Silicon Photomultiplier technology in gamma-ray astronomy and
achieving not only remote, but robotic operation of the telescope.

\section{Towards Robotic Operation}
After a brief period of on-site operation, the FACT Collaboration achieved complete remote operation of the telescope.
Two web interfaces are used to plan and execute data taking.
First, a scheduler to plan the observations\footnote{\url{https://www.fact-project.org/schedule}} and second, the smartfact interface\footnote{\url{http://fact-project.org/smartfact}},  a status and control interface. \citep{bretz2014nara}

The necessity for human interaction was reduced step by step.
It was achieved that shifters only have to startup the telescope in the evening, monitor the telescope and the environmental conditions during observation and shut down the telescope in the morning.
Human interaction during data taking is only needed in exceptional conditions like strong winds or heavy rain.
This automization lead to more than 2300 hours of data taken in the last twelve months.

\section{The \texttt{pycustos} library and the \texttt{shifthelper} program}
To reach the target of robotic operation, the need for humans to monitor the telescope system and the environmental conditions needed to be reduced even further.
For this purpose, the \texttt{shifthelper} program was created, written in \texttt{python}.
The \texttt{shifthelper} performs regular checks of the system status and environmental conditions and calls the shifter in case human intervention is needed via a webservice called \texttt{twilio}\footnote{\url{https://www.twilio.com}}.
The program can also send texts and images, e.\,g.\ plots, via the \texttt{Telegram}\footnote{\url{https://telegram.org}} messenger.

For about a year, the \texttt{shifthelper} was used by the shifters on their computers.
Currently, it is further developed to be a web service, automatically getting the information whom to notify in case human interaction is necessary from a database.A first version is being deployed and evaluated.

The infrastructure performing checks and notification was moved out into its own library called \texttt{pycustos}\footnote{\url{https://github.com/fact-project/pycustos}}.
More possibilities to send notifications have been added, including sending emails via \texttt{smtp}, posting alerts to a \texttt{REST}-interface via \texttt{http} and basic logging to a file.

Here is a basic example how you could use the custos library to automatically call the fire department in case your house is on fire:

First a \texttt{FireCheck} class is implemented, inheriting from the \texttt{pycustos} abstract base class \texttt{IntervalCheck}. 
The only method we need to override is the \texttt{check} method.
It creates a message via its \texttt{critical} method, in case our house is on fire.

\begin{verbatim}
from custos import Custos, IntervalCheck, TwilioNotifier
import my_house


class FireCheck(IntervalCheck):
    def check(self):
        if my_house.is_on_fire():
            self.critical('The house is on fire!')

if __name__ == '__main__':

    twilio_notifier = TwilioNotifier(
        sid='<your twilio sid>',
        auth_token='<your twilio token>',
        number='<your twilio number>',
        recipients=['112'],  # or '911'
    )

    fire_check = FireCheck(interval=30)

    with Custos(
            checks=[fire_check],
            notifiers=[twilio_notifier]
            ) as custos:
        custos.run()
\end{verbatim}

To place calls, an instance of the \texttt{TwilioNotifier} is created, which is given the necessary credentials to use the \texttt{twilio} service and list of recipients, for now just with one entry.
In the main program, the \texttt{FireCheck} is instantiated and given an interval of 30 seconds between subsequent checks.

Both checks and notifiers are bound together by the \texttt{Custos} class,
which runs as event loop after calling \texttt{custos.run}.
To allow more complex use cases, recipients of a notifier can either be a simple \texttt{list}, a \texttt{dictionary} mapping categories to lists of recipients or a function
that takes the category and returns a list of recipients.
Such a function is used to query the current shifter or the resposible expert in the \texttt{shifthelper} program.
An image can be attached to a message and notifiers able to send images, e.\,g.\ the \texttt{SMTPNotifier} and the \texttt{TelegramNotifier} will relay these images to the recipients.

\section{Additional tools}

\subsection{The \texttt{darkspot} program}
FACT is performing measurements of trigger rate vs.\ trigger threshold each night to monitor atmospheric properties.
These measurements are done pointing to a dark spot in the sky close to zenith.
To find such a spot and schedule the measurement the shifter had to select a suited pointing position by hand, e.\,g.\ using software like \texttt{stellarium}\footnote{\url{http://www.stellarium.org/}}.
To automatize this task, the \texttt{darkspot}\footnote{\url{https://github.com/fact-project/darkspot}} program was created.
Using the Hipparcos star catalogue~\citep{perryman1997hipparcos}, the area with the lowest light flux for the field of view of FACT is selected.

\subsection{The \texttt{la\_palma\_overview} program}

To get a quick overview about the conditions of a night, the \texttt{la\_palma\_overview}\footnote{\url{https://github.com/fact-project/la_palma_overview}} program was developed. 
Continuously merging the streams of several webcams on the Roque de los Muchachos, weather information and system status reports, it creates a timelapse video.

\section*{Acknowledgement} This work has been supported by the DFG, Collaborative Research Center
SFB 876, project C3 (\url{http://sfb876.tu-dortmund.de/}).

\bibliography{P6-19}

\end{document}